\documentclass[a4paper]{jpconf}
\usepackage{graphicx}
\usepackage{wrapfig}
\usepackage{amssymb}
\begin{document}

\title{Selected results from IceCube}

\author{Teresa Montaruli for the IceCube Collaboration}

\address{Département de Physique Nucléaire et Corpusculaire, Faculté de Sciences, Université de Genève, 1205 Genève}

\ead{teresa.montaruli@unige.ch}

\begin{abstract}
Neutrino astronomy saw its birth with the discovery by IceCube of a diffuse flux at energies above 60~TeV with intensity comparable to a predicted upper limit to the flux from extra-galactic sources of ultra-high energy cosmic rays (UHECRs). While such an upper limit corresponds to the case of calorimetric sources, in which cosmic rays lose all their energy into photo-pion production, the first statistically significant coincident observation between neutrinos and gamma rays was observed from a blazar of intriguing nature.
A very-high-energy muon event, of most probable neutrino energy of 290 TeV for an $E^{-2.13}$ spectrum, alerted other observatories triggering a large number of investigations in many bands of the electromagnetic (EM) spectrum. A high gamma-ray state from the blazar was revealed soon after the event and in a follow-up to about 40 days. A posteriori observations also in the optical and radio bands indicated a rise of the flux from the TXS 0506+056 blazar. A previous excess of events of the duration of more than 100~d was observed by IceCube with higher significance than the alert itself. These observations triggered more complex modeling than simple one-zone proton synchrotron models for proton acceleration in jets of active galactic nuclei (AGNs) and more observations across the EM spectrum.

A second piece of evidence was a steady excess of about 50 neutrino events with reconstructed soft spectrum in a sample of lower energy well-reconstructed muon events than the alert event. A hot spot was identified in a catalog of 110 gamma-ray intense emitters and starburst galaxies in a direction compatible with NGC 1068 with a significance of $2.9\sigma$. NGC 1068 hosts a mildly relativistic jet in a starburst galaxy, seen not from the jet direction but rather through the torus. This Seyfert II galaxy is at only 14.4~Mpc from the Earth.
The source turned out to be also the hottest spot of an all-sky search. Analyzed cumulatively, the catalog excess was $3.3\sigma$ with the contribution of NGC 1068 and TXS 0506+056, as expected, and the other 2 sources, PKS 1424+240, and GB6 J1542+6129, with similar features to TXS 0506+056, indicating that they might all be Flat Spectrum Radio Quasars (FSRQs).

While all these observations and the directions of the measured events contributing to diffuse fluxes hint to their extra-galactic origin, a few percent level contribution might be the end of a lower energy `granted' flux of neutrinos from interactions of cosmic rays in the Galactic Plane. This relevant observation is within the reach of IceCube and other neutrino telescopes.
These aspects were discussed at the conference and are summarised in this write-up.
\end{abstract}

\begin{wrapfigure}{L}{0.55\textwidth} 
  \centering
      \includegraphics[width=0.55\textwidth]{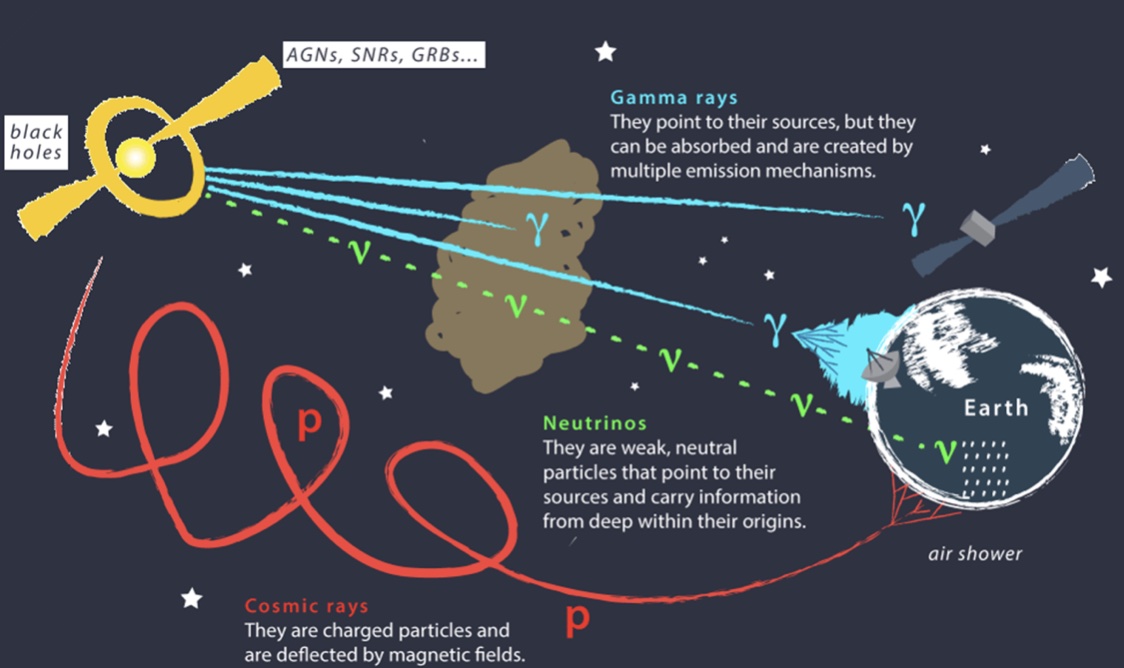} 
    \caption{\label{fig:MM} \small \it Multi-messengers from hadronic accelerators (credit J.A. Aguilar \& J. Yang).
      } 
\vspace{-15pt}
\end{wrapfigure}

\section{Multi-Messenger High-Energy Astrophysics}

The cosmic ray (CR) flux, studied since its discovery in 1912 by V. Hess, is the ``astroparticle'' component of cosmic radiation, composed also of photons from the radio to the gamma-ray band. 
The measurement of a CR extending from GeV energies up to about $10^{21}$~eV has triggered since then many questions. Nowadays the most compelling aspects concern their unknown origin, how they emerge out of their powerful accelerators and are accelerated, their diffusion and influence on the evolution of the Galaxy. 

Multi-messenger high-energy astrophysics is the extension of the multi-wavelength (MWL) exploration of the cosmos through multiple messengers with common origin (see Fig.~\ref{fig:MM}) with the potential to unravel the origin of CRs. CRs cover energies beyond the mass of the proton, hence they correspond to the gamma-ray band of photon radiation. Hence, three of the messengers, not including gravitational waves, are connected by the fact that gamma-rays and neutrinos are secondaries of proton/nuclei interactions on the matter or ambient radiation in sources, on the radiation that pervades the universe to us or on the Galactic plane. Neutrinos can extend the accessible horizon of gamma-rays to regions where they are absorbed. 
Despite their different cross section, and so different horizons and propagation properties, the coincident observation of a neutrinos and gamma rays is a strong indication of acceleration of protons/nuclei in a CR source.

\section{Neutrino Telescopes and IceCube}
\begin{wrapfigure}{L}{0.65\textwidth} 
  \centering
      \includegraphics[width=0.6\textwidth]{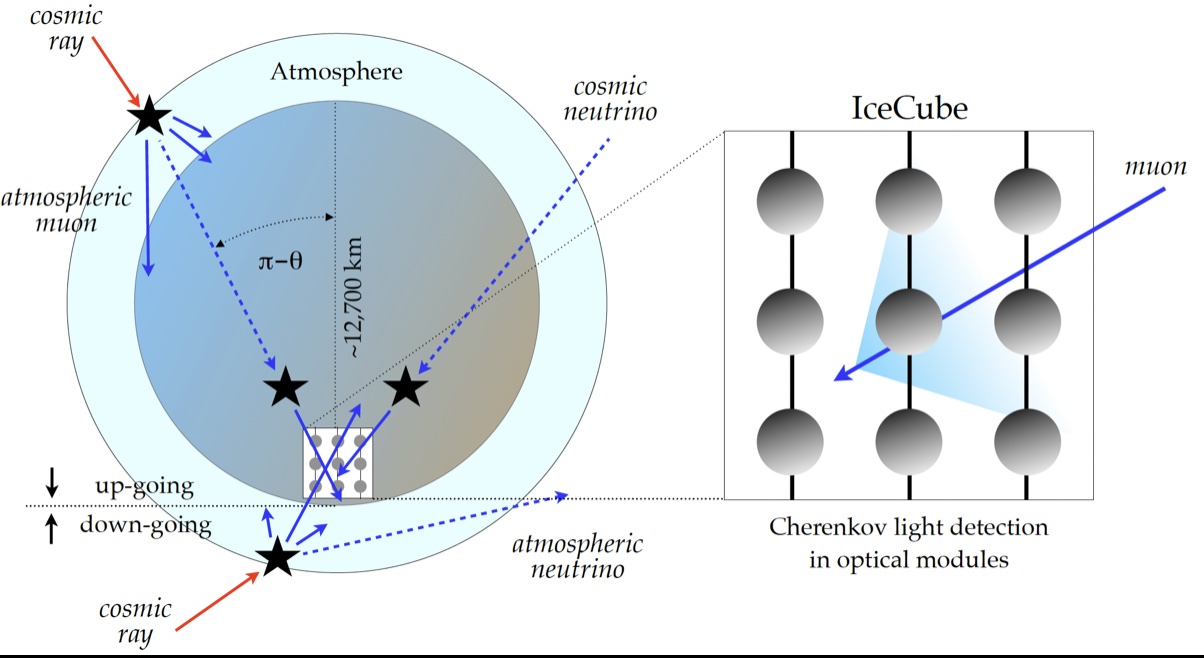} 
    \caption{\label{fig:principle} \small \it Principle of indirect $\nu$ detection in NTs.
      } 
\vspace{-10pt}
\end{wrapfigure}

A neutrino telescope (NT), for instance located at the South Pole as IceCube, is an array of synchronised (to ns level) photo-detectors measuring the time and energy of photons emitted by the charged secondaries of neutrinos through Cherenkov light emission (see Fig.~\ref{fig:principle}). As it happens for gamma-ray astronomy, neutrinos are indirectly detected as their properties are inferred from those of secondaries. Neutrinos might come directly from the cosmos or they are generated from cosmic ray (CR) interactions in the atmosphere that produce also muons, which above TeV energies are capable of penetrating to the detector depth also from quite inclined directions. 

Depending on the neutrino flavor, there are two main topologies of neutrino events: tracks from muon neutrino deep inelastic charge current interactions and cascades from other flavor neutrinos and neutral current interactions of all flavor neutrinos. Hints for ``double bump events'', characteristic of tau neutrinos above few PeV have been recently found \cite{IceCube:2015vkp}. As tau neutrino production in the atmosphere has a low probability at the observed high energies, they can be tracers of neutrino oscillations along cosmic baselines. Additionally, one cascade event has been measured possibly generated by an anti-electron neutrino event of about 6 PeV \cite{IceCube:2021rpz} in the Glashow resonance region (see also Fig.~\ref{fig:spectrum_mm}).
Typical rates in IceCube are 2.5~kHz for downgoing atmospheric muons, about $2 \times 10^5$ upgoing atmospheric $\nu$s/yr, and above 60~TeV  $\sim 10 \nu$s/yr, mostly of astrophysical origin.

\begin{wrapfigure}{L}{0.62\textwidth}
  \centering
      \includegraphics[width=0.62\textwidth]{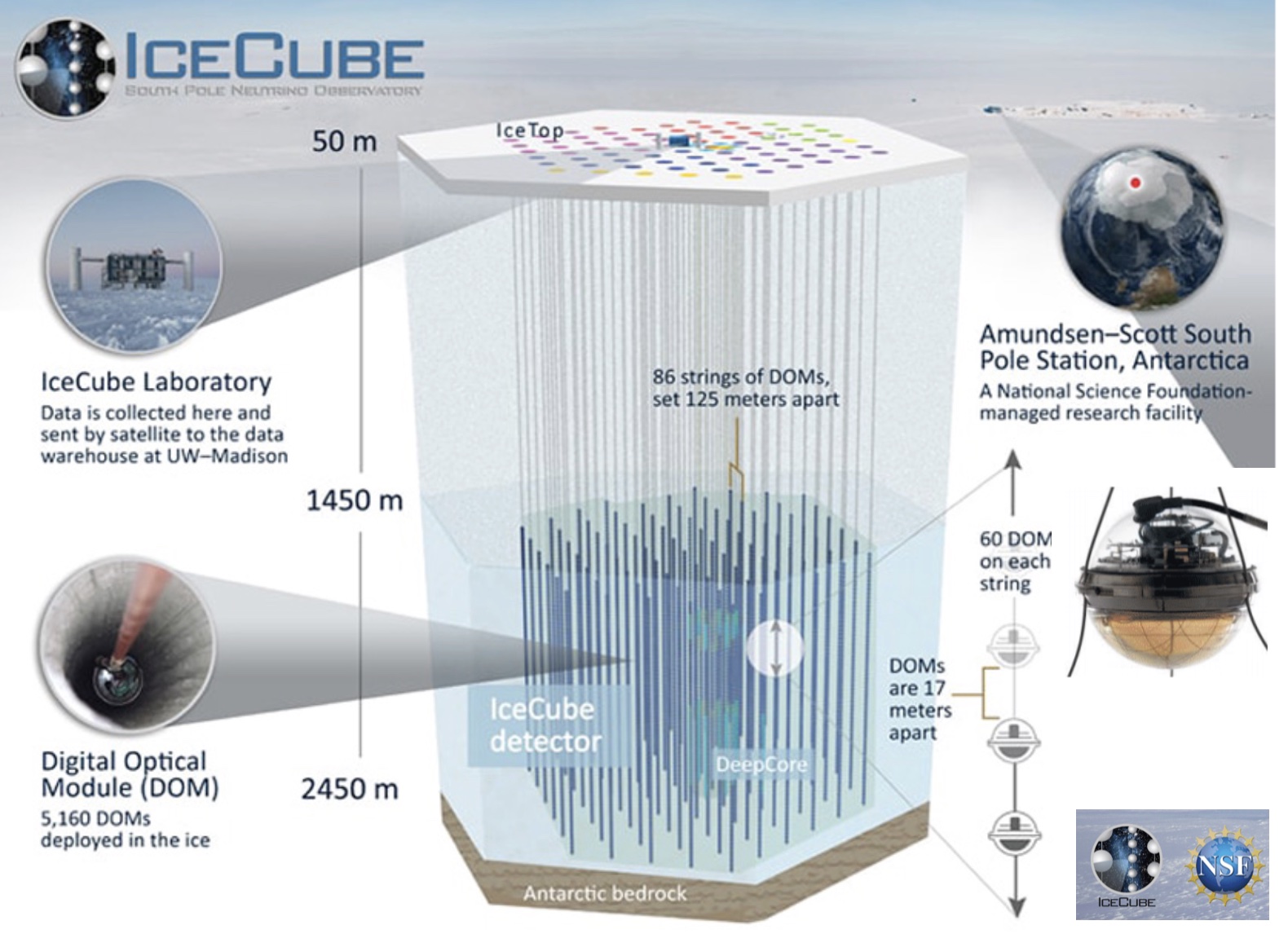} 
    \caption{\label{fig:detector} \small \it The IceCube in-ice detector, the IceTop surface array on top of the strings and downward-looking PMT in the DOM.
      } 
\vspace{-10pt}
\end{wrapfigure}

The about 5160 digital optical modules (DOMs) instrumenting IceCube are located between about 1450~m and 2450~m in the depths of the polar ice along 86 strings, separated by about 125~m and with vertical distance between DOMs of about 17 m (see Fig.~\ref{fig:detector}). On top of each string, there are 2 tanks composing the extensive air shower array IceTop.
Among the 86 strings, 8 are more densely instrumented and named DeepCore. They are dedicated to physics below 100~GeV, such as neutrino oscillations and dark matter studies. The strings of IceCube are cables carrying power and digitised data in the motherboard of the DOMs to the surface. The readout electronics of the downward-looking 10-inch photomultiplier (PMT) signals from Cherenkov light emitted by ultra-relativistic charged particles. This light, shown on the right of Fig.~\ref{fig:principle}, is emitted at an angle of $\sim 41^{\circ}$.

A random background of $\sim 1$~kHz is emitted by relativistic positrons in the beta decay of $^{40}K$ from the 0.5"-thick spherical glass pressure vessel that houses PMTs and the electronics \cite{IceCube:2016zyt}. Most of the PMTs have quantum efficiency (QE) of 25\% at $\sim 390$~nm, and about 7\% of the PMTs, mostly in DeepCore, have 34\% QE. 
The time resolution of PMTs is $\sim 2$~ns, when illuminated by narrow pulses, but the dispersion of photons due to scattering in the ice is the limiting factor for the reconstruction~\cite{IceCube:2010dpc}. Recently, the modeling of the light propagation properties of the ice has been extremely improved, with the addition of an asymmetric light diffusion effect in the birefringent poly-cristalline  micro-structure of the ice~\cite{2021JInst..16C9014R} enabling better angular resolution also for the cascade channel. Reconstruction techniques are also advancing by deep learning employment, such as Graphic Neural Networks running on GPUs \cite{IceCube:2010dpc}. Another recent improvement concerns the charge extraction which has been recently carefully re-calibrated \cite{2020JInst..15P6032A} and data/simulation have been reprocessed with a more correct representation of the Single Photoelectron pulse.

\begin{wrapfigure}{R}{0.5\textwidth} 
  \centering
      \includegraphics[width=0.5\textwidth]{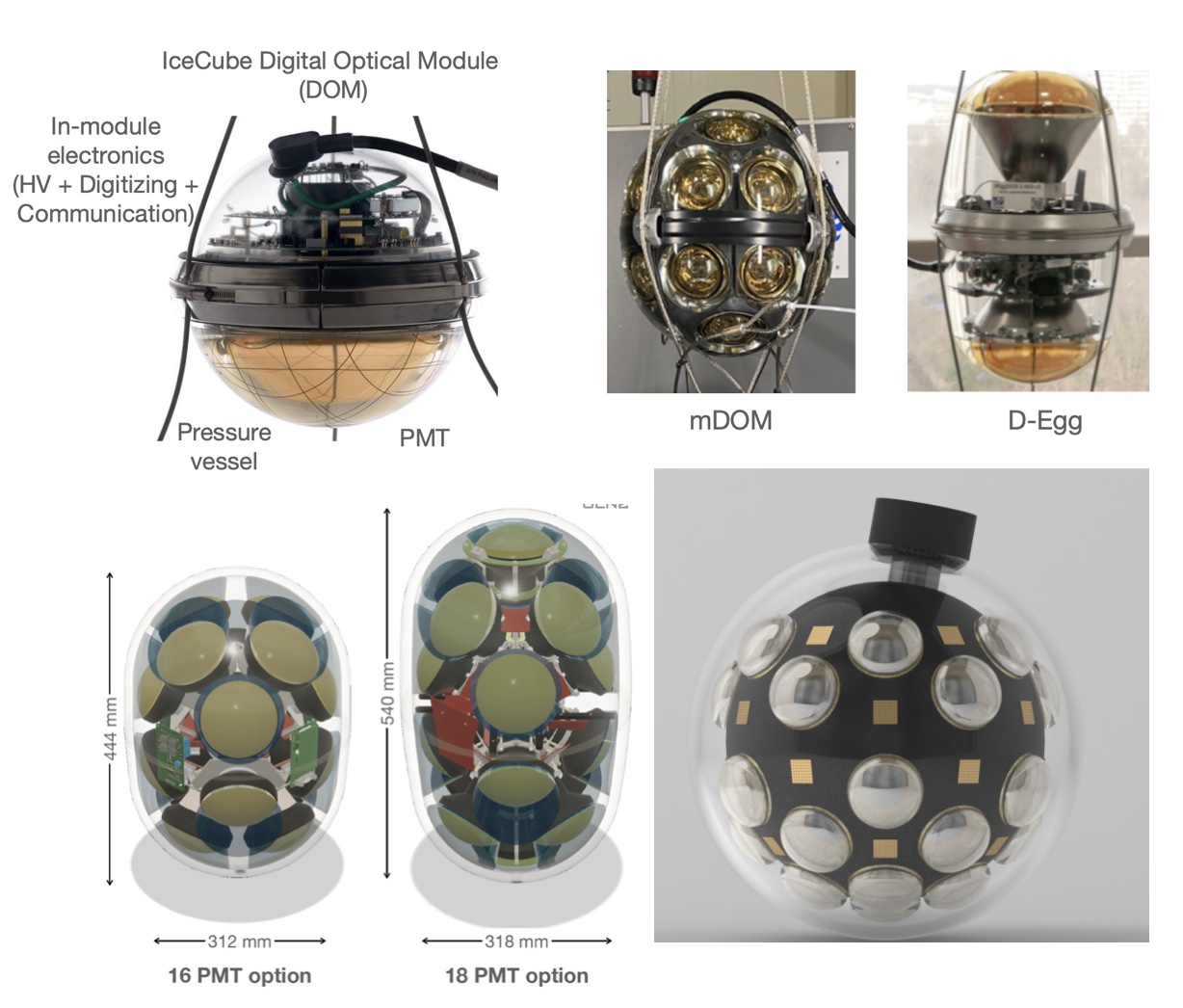} 
    \caption{\label{fig:Photosensors} \small \it
      IceCube DOM, D-Egg and mDOM prototypes for the IceCube Upgrade and the designs for IceCube-Gen2, all based on PMTs; the concept proposed by TRIDENT including SiPMs~\cite{Ye:2022vbk}.} 
\vspace{-10pt}
\end{wrapfigure}

The future of IceCube has started: the upgrade of IceCube, namely the addition of 7 dense strings with 700 optical modules followed by IceCube-Gen2. Its science scope is described in \cite{2021JPhG...48f0501A}. It expands the scope of IceCube by enabling better statistics to pin down the origin of the high-energy diffuse flux and better constraining parameters of models that determine particle acceleration and propagation through cosmic media,  also through better flavor measurements and intense multi-messenger programs; addressing neutrino properties, such as the existence of neutrino sterile and mixing. IceCube-Gen2 will increase the annual rate of observed cosmic neutrinos by a factor of ten compared to IceCube and will be able to detect sources five times fainter than its predecessor. It will be about a factor of 8 bigger than IceCube with additional 120 strings with 9600 optical modules and a surface array. 
The evolution of the optical modules is shown in Fig.~\ref{fig:Photosensors}, where the compromise proposal for IceCube-Gen2, resulting from the experience based on two developed prototypes, is shown. The design inherits from the KM3NeT concept and profits of better uniformity and smaller transit time spread of small PMTs compared to larger ones. 

Traditionally, NTs and many other Cherenkov detectors in astroparticle are based on very advanced PMT technology. 
Recently, gamma-ray astronomy has pioneered also SiPMs, which might become an option for the future of NTs as well.
A recently proposed concept for a neutrino telescope design in the Pacific is also shown in the picture, combining the large surface coverage of PMTs with the fast rise time properties of silicon photomultipliers~\cite{2022icrc.confE1043H,Ye:2022vbk}. 
Montecarlo simulations indicate an improvement of about 40\% in the angular resolution.

\begin{wrapfigure}{R}{0.56\textwidth} 
  \centering
      \includegraphics[width=0.55\textwidth]{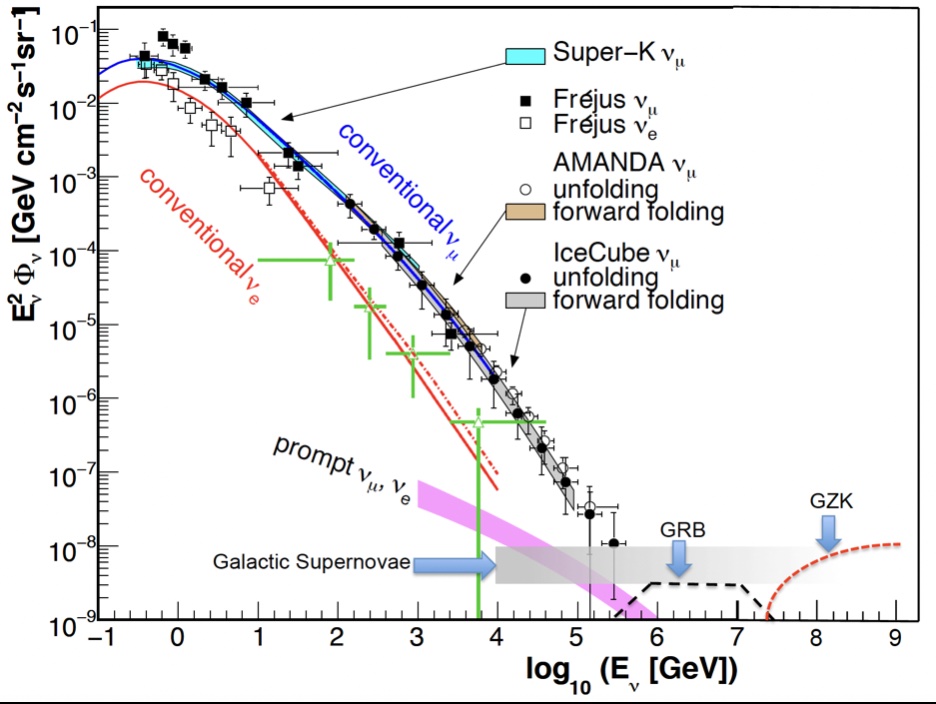} 
    \caption{\label{fig:atmo} \small \it
      The measured atmospheric $\nu$ flux by IceCube for the muon and electron components. The prompt and cosmic $\nu$ fluxes are indicated for illustration.
      } 
\vspace{-10pt}
\end{wrapfigure}

New samples based on improved cascade reconstructions will become more and more important for neutrino astronomy, as it is understandable from Fig.~\ref{fig:atmo}. They profit from excellent energy resolution with respect to muon samples (order of 10\% compared to about of a factor of 2 in IceCube) but angular resolutions which can be pushed to $5^{\circ}$ level. The electron neutrino flux from the atmosphere is about an order of magnitude lower than for muon neutrinos allowing to lower the energy threshold for the astrophysics neutrino searches, in a region where fluxes are higher. Nonetheless, the muon channel remains the only one with $\sim 0.2^{\circ}$ angular resolution above 10~TeV \cite{IceCube:2019cia}.
Atmospheric neutrinos, are for NTs also a signal for many studies, such as neutrino oscillations and hadronic models in the atmosphere. The prompt neutrino flux, visible in the figure, is harder as it comes from the prompt decay of charmed mesons. It is not yet measured by IceCube. The atmospheric neutrino beam in the figure is compared to expectations for cosmic fluxes and the gray region represents the upper bound of diffuse extra-galactic fluxes by Waxman \& Bahcall \cite{1998PhRvD..59b3002W}, discussed in Sec.~\ref{sec:grantedfluxes}.

\section{The diffuse astrophysical $\nu$ fluxes}
\label{sec:grantedfluxes}

The diffuse neutrino fluxes with energy $> 60$~TeV from 7.5~yr of high-energy starting events \cite{IceCube:2020wum}, dominated by downgoing cascades, and 9.5~yr of muon neutrinos \cite{IceCube:2021uhz}, shown in Fig.~\ref{fig:spectrum_mm}, represent solid evidence for astrophysical $\nu$s.
Such evidence is confirmed by ANTARES and GVD excess hints in diffuse fluxes with respect to atmospheric $\nu$s, and will be soon firmly detected by both GVD and KM3NeT. 

The comparable energy rate densities (see Fig.~\ref{fig:spectrum_mm}) of UHECRs, measured by the Pierre Auger and Telescope Array experiments \cite{Abreu_2021,Ivanov:2021mkn}, and the $>50$~TeV–PeV neutrinos discovered by IceCube~\cite{Aartsen2013,2020arXiv200109520I}, can be explained by a {\bf unified origin} assuming photo-meson interactions, e.g. extragalactic sources. 
The neutrino flux is at the level of 
the upper limit on extragalactic neutrino fluxes predicted by Waxman and Bahcall in 1998 \cite{1998PhRvD..59b3002W}. This seems to hint at calorimetric sources as possible contributors to the flux.

As discussed in \cite{PhysRevD.102.083023} and was discussed at the conference by S.~Yoshida, for a source of joint UHECR and neutrinos, the cosmic-ray acceleration time scale must be faster than the dynamical time scale, 
which depends on the Lorentz factor and the accelerating region dimension; UHECRs must be accelerated before they lose energy, e.g. due to synchrotron; UHECRs can escape their sources before losing their energies, hence dynamical time must be faster than the synchrotron cooling time. This results in an upper limit on the optical depth of photo-pion interactions ($\tau_{p\gamma}$) and boundary conditions to the magnetic field in the plasma reference frame. For this condition to happen, the system should not be calorimetric to imply UHECR and neutrino observation, but rather 'optically thin' \cite{PhysRevD.102.083023,PhysRevD.64.023002}.

The Waxman and Bahcall upper bound or {\bf calorimetric limit} is obtained when all the proton energy goes into pion production (interaction length $\tau_{p\gamma} =1$). 
Its value is $E^2 \times \frac{dN}{dE} \sim 10^{-8}$ GeV s$^{-1}$ sr$^{-1}$ cm$^{-2}$. It constrains jet models and also proton-proton assumptions (as the interaction length is not larger for $pp$ than $p\gamma$). Higher $\nu$ fluxes can only be produced in hidden-core AGNs or opaque source regions with $\tau_{p\gamma} >> 1$ from which only neutrinos escape. This upper bound was calculated in the assumption that UHECRs are dominantly protons and for an $E^{-2}$ flux fitted from above $10^{19}$~eV and extrapolated to lower energies where the energy spectrum might be different. Hence, extending this condition in the region of $\sim 100$~TeV neutrinos requires strong assumptions~\cite{Mannheim:1998wp}.

A joint study between NTs and UHECR experiments excluded possible correlations between UHECR directions and neutrinos \cite{IceCube:2022osb,ANTARES:2019ufk}. This could also be due to the different horizon of the two messengers and samples of neutrinos dominated by sources beyond the horizon of UHECRs.

This flux is at lower energy and more intense than what is expected also from cosmogenic neutrinos from interactions of cosmic rays with cosmic microwave radiation on the way to us, recently constrained by IceCube \cite{2018arXiv180701820I}. Hence, it remains open the question:
{\em Which are the sources contributing to the measured diffuse IceCube neutrino flux?}.

\subsection{The Galactic Plane fluxes}
Based on the direction of the contributing measured events, the fraction of the diffuse IceCube neutrino flux at $\sim 60$~TeV in the direction of the Galactic Plane is less than 10\% \cite{IceCube:2020wum}. 
Galactic cosmic rays wandering millions of years in the Galactic Plane are expected to contribute neutrino and gamma-ray fluxes in interactions with interstellar matter.

Recent works have shown the deep link between the observed gamma-ray diffuse flux from the galactic plane region and the galactic cosmic ray measured component spectra, and produced expected neutrino fluxes~\cite{Ahlers:2015moa,2015ApJ...815L..25G}. Predicted neutrino fluxes in Ref.~\cite{2015ApJ...815L..25G} have been constrained by IceCube \cite{2017ApJ...849...67A} and IceCube jointly with ANTARES \cite{ANTARES:2018nyb}, limiting the expected contribution from the Galaxy to the diffuse flux $>60$~TeV to a few percent and indicating that this galactic flux is at a reach. 

 It was also discussed the relevance of ultra-high energy gamma-ray measurements (e.g. from Tibet AS$\gamma$ \cite{TibetASgamma:2021tpz} and future expected ones by LHAASO), as the CR proton spectrum is still not perfectly known in the region of the knee~\cite{2022A&A...661A..72B,PhysRevD.98.043003}. This is due to some discrepancy between KASCADE and IceTop proton knees at about 4~PeV, and to some extent to the discrepancies in the helium knee at about $Z \times 4$~PeV, where Z is the number of protons in the ionized nucleus. 

The knowledge of the fluxes of all components of the galactic region of the CR spectrum extending to the multi-PeV major bending region, called the knee, is not only relevant for the predictions of the diffuse galactic neutrino fluxes but it is also of fundamental relevance for indirect detection of messengers from the ground (atmospheric neutrinos and muons in NTs and cosmic ray shower backgrounds for gamma-ray experiments). 
Prediction of this backgrounds through CORSIKA simulation \cite{Heck:1998vt,2021arXiv211211761A} requires as input the spectral shape of the various components of the CR flux, as well as hadronic models, which are still affected by energy-dependent uncertainties.

\subsection{Jetted AGNs}

An upper limit was set by IceCube on gamma-ray emitting blazars in the 2LAC catalog of Fermi~\cite{2017ApJ...835...45A}, dominated by Flat Spectrum Radio Quasars (FSRQs) and including BL Lacs (describe below). The neutrino diffuse flux might be explained by these blazars up to 27\% for $E^{-2.2}$ and 50\% for $E^{-2.5}$. Despite these limits make the strong assumption of a common spectral shape of all AGNs, it is reasonable to assume that another class of sources might contribute to the diffuse neutrino flux.

A few compelling observations on blazars exist by now, possibly hinting at a particularly powerful class of blazars, which might or not have coincident flares in the $\gamma$-ray band \cite{Halzen_flare,2022arXiv220200694H}.
One of them concerns the 270 TeV neutrino alert sent by IceCube on Sep. 2017 to the astronomical community \cite{IceCube:2018dnn}. {\it Fermi}-LAT, as well as MAGIC follow-up observations~\cite{MAGIC:2018sak}, confirmed the presence inside the error region of about $0.5^\circ$ from the IceCube event direction of a flaring blazar, TXS 0506+056, located at redshift of 0.33 with a chance probability of 3$\sigma$.

The picture appeared immediately complex for TXS 0506+056, previously classified as high-frequency peaked BL Lacertae (HBL BL Lac) \footnote{HBL and low-frequency peaked LBL are the blazars with the frequency of the synchrotron bump of the SED  $>10^{15}$~Hz or $<10^{14}$~Hz, respectively.}, possibly being a `masquerading' FSRQ.  FSRQs are relatively luminous LBL blazars with strong, optical-UV emission lines in addition to the non-thermal continuum~\cite{2019MNRAS.484L.104P,MAGIC:2018sak}, a signature of the Broad Line Region (BLR) outside the jet, likely photo-excited by a radiatively efficient accretion disk around a supermassive black hole (SMBH). Masquerading FSRQ would dissimulate BL Lacs as broad lines are not clearly visible due to non-thermal jet emission \cite{2019MNRAS.484L.104P}.
Typical leptonic models for BL Lacs are Synchrotron Self Compton models (SSC), where emitted synchrotron photons up-scatter by Inverse-Compton (IC) on the same emitting lepton population (hence named single zone models), while, for FSRQ, IC can happen on external fields provided by the thermal photons outside of the jet (External Inverse Compton models - EIC). 

The MWL Spectral Emission Distribution (SED), including the 2017 flare, cannot be explained by pure hadronic models where gamma-ray emission is dominated by proton synchrotron \cite{2001APh....15..121M}. Single-zone models can barely accommodate the 2017 flare, but predict low rates of neutrinos. These can be increased ($\sim 0.1 \nu/\rm yr$) in lepto-hadronic models, but they have demanding energetics for protons the jet ($> 10^{46 - 47}$~erg/s), at times super-Eddington~\cite{2018ApJ...864...84K,2019NatAs...3...88G,Cerruti:2018tmc}. Solutions with proton synchrotron on external photon fields, e.g. on thermal radiation \cite{2018ApJ...864...84K}, radiatively inefficient accretion flow (RIAF) \cite{2019MNRAS.483L.127R} and layered jet \cite{MAGIC:2018sak} are less demanding on the energetics, but do not fit the second more significant flare of about 100 d between 2014-2015 observed in the historical data of IceCube with the significance of $3.5\sigma$\cite{IceCube:2018cha}.

\begin{wrapfigure}{L}{0.66\textwidth} 
  \centering
\includegraphics[width=0.66\textwidth]{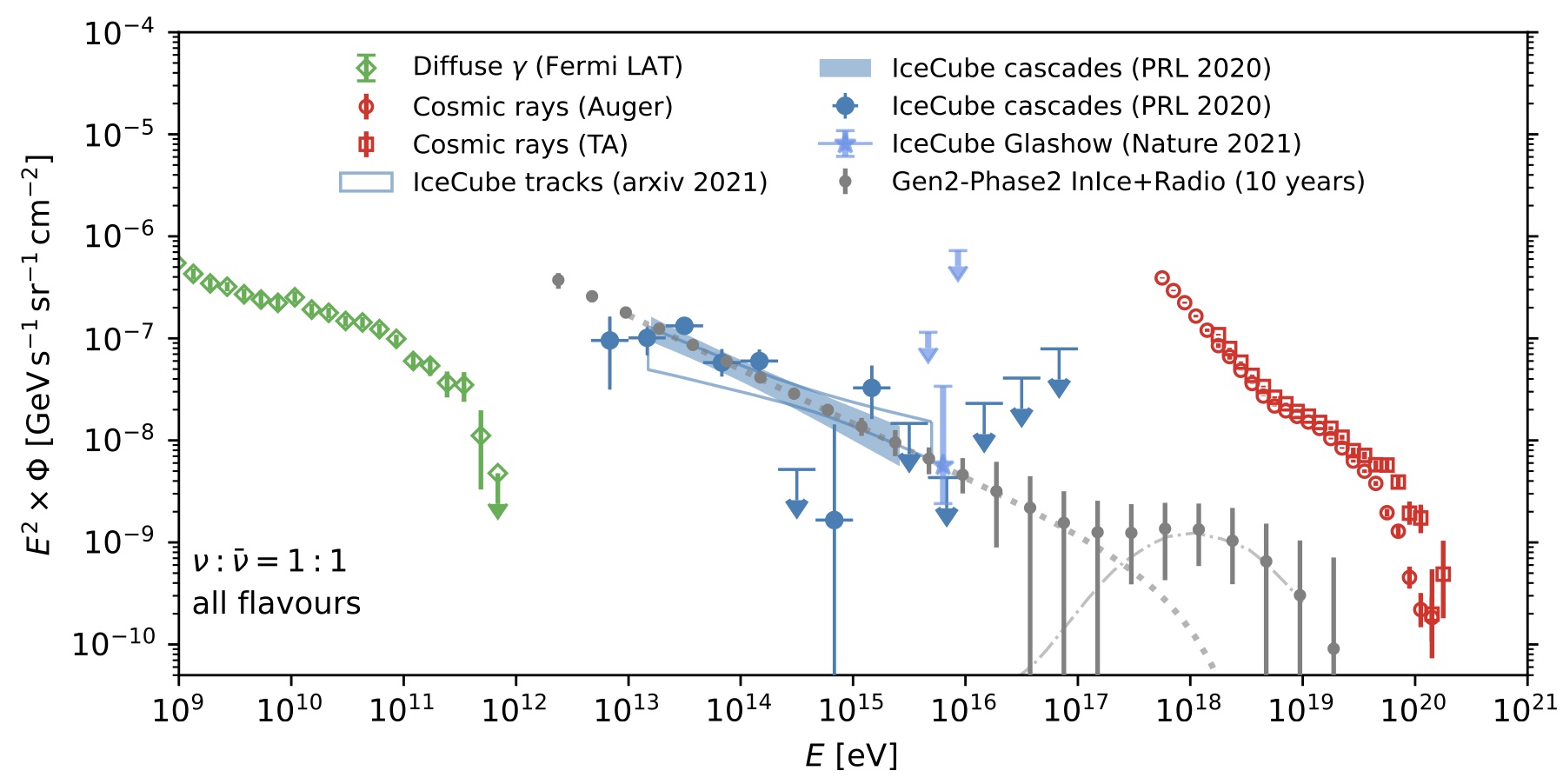}
\caption{\label{fig:spectrum_mm} \small \it The UHECR diffuse spectrum \cite{Abreu_2021,Ivanov:2021mkn}; the isotropic diffuse $\gamma$-rays measured by {\it Fermi}-LAT \cite{Fermi-LAT:2014ryhy}; the  high-energy $\nu$ starting events of IceCube \cite{IceCube:2020wum,Aartsen2013} (mainly cascades) and the 9.5 yr-sample of up-going tracks \cite{IceCube:2021uhz}. The star is from the Glashow resonance cascade event \cite{Glashow2021}. The prediction of the diffuse flux in 10 yr of IceCube-Gen2 data is shown in gray with 68\% C.L.  bars~\cite{2022arXiv220211120V}. Plot in \cite{2022arXiv220308096A}.
} 
\vspace{-10pt}    
\end{wrapfigure}
 
Single zone models do not fit the higher neutrino flux implied by this observation as the larger amount of photons from pions in the emitting region is in tension with the measured Swift X-ray flux \cite{2019ApJ...874L..29R}. This can be shifted to MeV energy and possibly imply an opaque region in photons emitting neutrinos physically separated, hence incompatible with single-zone models.

While both TXS 0506+056 flares do not have the significance of discovery, some interesting observations were made in coincidence with the 2017 and 2014-2015 observations by IceCube. E.g. the MASTER global optical telescope network reported that TXS 0506+056 was in a quiet state 73~s after the IceCube 2017 event, but 2~hr after it the optical flux increased at $50\sigma$ level on top of the background, namely the biggest variation they recorded since 2005 \cite{2020ApJ...896L..19L}. Additionally, during the 100~d neutrino-flare period across 2014-2015, another longer flare with about 3.5$\sigma$ significance was observed.
VLBA 15~GHz observations in 2009-18 on the structure of the jet of TXS 0506+056, showed a curved jet structure, possibly a precessing jet with 10~yr period, with the 2017 alert in the bright precession phase or a cosmic collider of 2 jets~\cite{2019A&A...630A.103B}. 
Additionally, in Ref.~\cite{2020A&A...633L...1R} it was noted that the 43~GHz radio VLBI observations between Nov.~2017 and May~2018 indicate a compact core with highly-collimated jet and a downstream jet showing a wider opening angle (slower) external sheath. 
The slower flow could serve as seed photons for interactions producing neutrinos.
Such a scenario, also advocated in \cite{MAGIC:2018sak}, is assumed in spine-sheath models of neutrino production \cite{2014ApJ...793L..18T,2018ApJ...864...84K}.

In summary, the TXS 0506+056 evidence revealed the option that multi-messenger programs have the potential to explore the structure of the acceleration mechanism and of the jet \cite{Petropoulou_2020}. Further observations of MAGIC and VERITAS \cite{MAGIC:2022gyf} in 2017-19, resulted in another flare with 4$\sigma$ significance in Dec. 2018 detected by MAGIC in 74 hr of observations, which could have been too short to produce evidence in IceCube.
Additionally, OVRO radio data indicate a rising flux. This demands further observations in neutrinos.

There are a number of other hints that are not significant in IceCube but have triggered some literature, and particularly animate discussions on what could be the class of blazars that with increasing statistics will produce signals in IceCube and its successor IceCube-Gen2.

Another golden alert of IceCube, a $\nu$ event of $\sim 300$~TeV~\cite{Stein},  was compatible with the location of PKS 1502+106. This is a high-redshift ($z\sim 1.8$) LBL and highly polarized FSRQ \cite{PKS_Fermi}. No gamma-ray flare was identified but rather an increase of the radio flux measured by the 40~m telescope OVRO \cite{Kun}. This is similar to the 2014-2015 flare of TXS 0506+056
during which the radio flare was increasing, as during the 2017 event.
A similar scenario of jet precession as for TXS 0506+056 was identified in the radio data by \cite{Britzen2021}: a precessing curved jet interacting with NLR clouds at a distance of 330 pc. A ring-like and arc-like configuration develop right before the neutrino emission, not present at all times. 

It was speculated in \cite{Halzen_flare,Kun} that TXS 0506+056, PKS 1502+106, and even PKS B1424-418 (observed with one of the biggest cascade neutrino events ever observed by IceCube named Big Bird) belong to a sub-class of $\sim 5\%$ blazars able to explain the IceCube diffuse flux. They could as well be efficient neutrino emitters while they are inefficient gamma emitters. 
 This would explain why the IceCube's already mentioned cumulative limits on blazars~\cite{2017ApJ...835...45A}  can be evaded.  

\subsection{Calorimetric sources}

Starburst Galaxies (SBGs) are characterized by a high formation rate (SFR), typically ranging from 10 to 1000~$M_\odot$/yr, proportional to the IR emission. Hence, SNe and star winds can be very efficient CRs factories and accelerators.
The SFR is also correlated to the radio/gamma-ray luminosity, as it is proportional to the number of acceleration sites such as supernova \cite{Kornecki:2021iuh,2021MNRAS.503.4032A}. 
These are sub-kpc regions, called Starburst Nuclei (SBNi), with gas density $>10^2$~cm$^{-3}$, IR emission $> 10^3$~eV cm$^{-3}$ and $B \gtrsim 10^2 \mu$G (indicating a high level of turbulence). They are acceleration sites of CRs, as well as the wind structures departing from them. Wind bubbles can accelerate CRs in a standard diffusive shock acceleration scenario to $\sim 100$~PeV for hundreds Myr\cite{Peretti:2021yhc}. 
In this scenario, it was found that SBG can explain the UHECR and diffuse $\nu$ flux~\cite{Condorelli:2022vfa}.

Some SBGs are identified as Syfert galaxies, making them intriguing objects where the core may host a hidden jet. 
While blazars shoot their jets against us, Seyfert galaxies (about 10\% of all galaxies) are seen across the torus (Seyfert II), the accreted matter surrounding the SMBH, or through the narrow (NLR) and BLR (Seyfert I). Broadening is related to the Doppler shift, so to the temperature of the gas heated in the inner part by accretion emission (NRL) and for the BLR also to velocity dispersion and orbital spin around the SMBH. 

Some years ago, the analyses in Ref.s~\cite{PhysRevD.88.121301,2014JCAP...09..043T} have pointed out that SFGs and SBGs cannot be the dominant sources of the whole $> 60$~TeV diffuse $\nu$ flux, without exceeding the non-blazar component of the {\it Fermi}-LAT extragalactic gamma-ray background (EGB) \cite{Fermi-LAT:2014ryhy} in the low energy part of Fig.~\ref{fig:spectrum_mm}. In another model on the wind bubbles, it was found that SBG can explain the UHECR and diffuse $\nu$ flux \cite{Peretti:2021yhc}.

\begin{wrapfigure}{L}{0.63\textwidth} 
  \centering
      \includegraphics[width=0.61\textwidth]{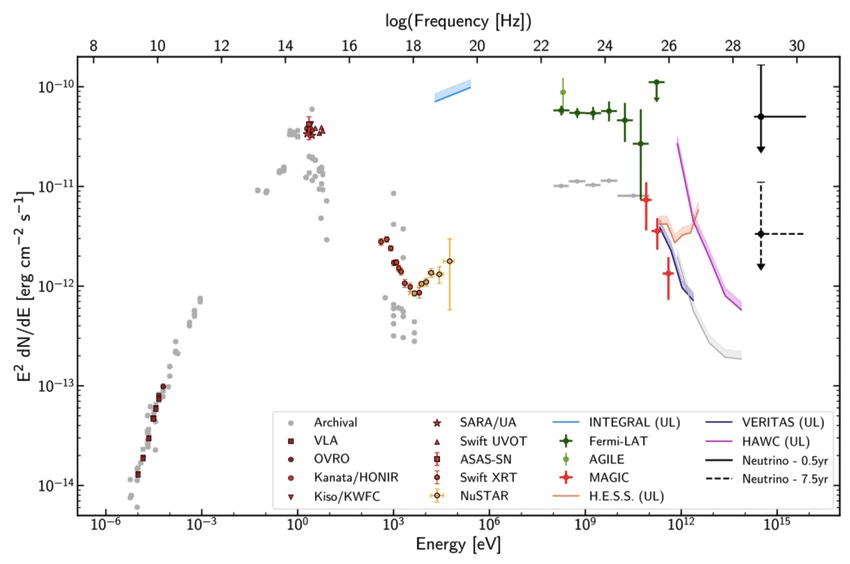} 
\caption{\label{fig:spectrum} \small \it
    SED for the blazar TXS 0506+056, based on observations obtained within 14 d of IceCube-170922A. Archival observations are in gray. The $\nu_{\mu}+\nu_{\mu}$ flux upper limits, that produce on average one detection like IceCube-170922A over a period of 0.5 (solid black error lines) and 7.5~yr (dashed black error lines) assuming an $E^{-2}$ spectrum at the most probable neutrino energy of $\sim 300$~TeV~\cite{IceCube:2018cha}.
      } 
\vspace{-10pt}
\end{wrapfigure}

Recently, the Seyfert II SBG NGC 1068 emerged as the hottest source in a catalog of 110 gamma-ray emitters sources and 8 SBGs, and the hottest spot in the full scanned skymap despite the large trial factor. 
About 50 signal-like events were fit with a reconstructed spectrum for a single power-law hypothesis of about $E^{-3.2}$.  
A cumulative emission population study of the catalog produced evidence of an excess of $3.3\sigma$ with respect to the atmospheric background. It is dominated by NGC 1068, which has a post-trial significance of $2.9\sigma$ among the background dominantly of atmospheric neutrinos. 
Interestingly, all other sources contributing principally to this $3.3\sigma$ excess were peculiar blazars, probably masquerading FSRQ \cite{Sahakyan:2022nbz}, namely TXS 0506+056, PKS 1424+240, and GB6 J1542+6129. TXS 0506+056  was expected from the former observation of the 2017 alert and second flare.

NGC 1068 hosts a mildly relativistic jet embedded in a Seyfert II Galaxy seen through its accretion disk. 
Such an accelerator is in the first place interesting because it is very close to us, and probably driven by different mechanisms than jet acceleration. Rather acceleration in wind bubbles and proton-proton interactions might be in place in SBG, similar to SFR in the Galaxy as the Cygnus cocoon or the Fermi Bubbles.

Interestingly, a significance of $4\sigma$ has been achieved for a small-scale anisotropy in the directions of starburst galaxies for a UHECR sample of energy $>45$~EeV  by  Pierre Auger~\cite{PierreAuger:2021rfz}.
In Ref.~\cite{2021PhRvD.104h3013A}, it is noted that one of the highest-energy $\nu$ event of IceCube, falls close to a UHECR anisotropy region with NGC 4945, as detected by Pierre Auger with a sample of UHECRs above 56~EeV collected in 3.5~yr \cite{2007Sci...318..938P}. This event is not statistically significant but may hint at the need for further observations.

 In \cite{2021MNRAS.503.4032A}, it is shown that using the measured spectra of 13 SFGs selected in Ref.~\cite{2020ApJ...894...88A}, and accounting for additional contributions from blazars, SFGs can have a relevant role in the diffuse flux observed by IceCube. 
In \cite{2020ApJ...894...88A}, the 13 candidates were found among a sample of 683 SFGs, with significance $>4.6\sigma$ in 10 yr of {\it Fermi}-LAT data between 100~MeV and 820~GeV showing. 
While the more intense Arp220 and Arp229 between them are at 100 million light-year distances, NGC 1068 is at about 46 million light-years and NGC 4945 is even closer, 13 Mly! 
After only 1.5 yr of {\it Fermi}-LAT data, it was pointed out that NGC 1068 and NGC 4945 are the first Seyfert II galaxies with high-energy $\gamma$-ray emission \cite{2010A&A...524A..72L}. 
Gamma-ray observations are available up to TeV energies only for NGC 253, detected by H.E.S.S. \cite{2005A&A...442..177A}, and M82 by VERITAS \cite{2009Natur.462..770V}. 

Additionally, a group of researchers found that another blazar PKS 0735+178 is just outside the localization error of $\sim 13^\circ$ at 90\% c.l. of the golden alert IceCube-211208A on Dec.~21, 2021 of most probable estimated energy of 172~TeV\cite{Sahakyan:2022nbz}. 
It was noted the coincidence that a Baikal-GVD event of most probable estimated energy of 43 TeV was detected about 4 hr after IceCube-211208A and inside the $5.5^\circ$ (50\% c.l. statistical error region)  from \cite{2021ATel15112....1D}. In \cite{Sahakyan:2022nbz}, it is remarked that the source doubled its X-ray flux on Dec. 17, 2021, less than $5 \times 10^3$ s of the IceCube event, and was in a very high state in $\gamma$-rays, UV and optical and other observations are quoted by Baksan and KM3NeT (and references therein).
This flaring source is of the class considered in
\cite{Plavin:2020emb,Plavin:2020mkf}. These AGNs are selected through VLBI data with strong parsec-scale cores and radio flares are searched for to exploit possible coincidence with neutrinos. A new catalog of more than 4000 sources is in \cite{Koryukova:2022abw}. While the authors have claimed a statistical excess, it is still a matter of debate worth further investigation. ANTARES has found an excess when using this catalog \cite{Illuminati:2021ezq}, which was not confirmed by IceCube. 

In summary, I showed that there are a few solid observations on potential neutrino sources, and many hints, that indicate how flourishing is this start of neutrino astronomy, and how much it needs multi-messenger and multi-observatory observations.

\subsection{Acknowledgments}
I wish to thank the Organisers of this conference, and particularly A. Insolia, F. Guarino and C. D'Aramo,  for the CRIS recurrent appointment of outstanding value.

\section*{References}
\bibliographystyle{iopart-num}
\bibliography{multimessenger}

\end{document}